\RequirePackage{fix-cm}
\documentclass[twocolumn,final]{svjour3}

\smartqed
\usepackage[pdftex]{graphicx}
\usepackage{etex}
\journalname{CGI2012} 
\usepackage{amsmath}
\usepackage{tikz}
\usetikzlibrary{shapes,arrows}
\usepackage{url}
\usepackage{listings}
\usepackage{color}
\usepackage{program}
\usepackage{balance}
\usepackage{cite}
\usepackage[caption=true,font=footnotesize]{subfig}

\begin{document}
%
\title{Real-time Animation Of Human Characters With Fuzzy Controllers}


\author{Koen Samyn \and Sofie Van Hoecke \and Bart Pieters \and Charles Hollemeersch \and Aljosha Demeulemeester
\and Rik van de Walle}
\institute{
K. Samyn \at Ghent University -– IBBT, ELIS Department, Ghent, Belgium,\\ \email{koen.samyn@ugent.be}
\and 
K. Samyn \at Howest University College, Kortrijk, Belgium
\and  
S. Van Hoecke \at Howest University College, Kortrijk, Belgium,  \\ \email{sofie.van.hoecke@howest.be}
\and 
S. Van Hoecke \and B. Pieters \and C. Hollemeersch \\ \and A. Demeulemeester \and R. van de Walle \at Ghent University -– IBBT, ELIS Department, Ghent, Belgium
}

\date{ }
\maketitle

\begin{abstract}
The production of animation is a resource intensive process in game companies. Therefore, techniques to synthesize animations have been developed. However, these procedural techniques offer limited \linebreak adaptability by animation artists. In order to solve this, a fuzzy neural network model of the animation is proposed, where the parameters can be tuned either by machine learning techniques that use motion capture data as training data or by the animation artist himself. This paper illustrates how this real time procedural animation system can be developed, taking the human gait on flat terrain and inclined surfaces as example. Currently, the parametric model is capable of synthesizing animations for various limb sizes and step sizes.
\end{abstract}

\section{Introduction}
Animation of human and non-human characters is a very important topic for the production of video games, movies or commercials. With the push for 3D worlds that are immersive and fully interactive, a large percentage of production time is spent on the development of character animation. Game studios commonly use handcrafted animations by animation artists or motion capture during production. However, because both techniques are very labor intensive, different algorithms to create procedural animation (both real-time and non real-time) have been proposed.

While talking to established animation artists and reading the standard work\cite{williams2009} for animation artists, it became clear that there is a large disconnect between the procedural animation community and the animation artists community. This disconnect is mainly situated in the conflicting goals of both communities. The artist community wants to retain full artistic control over the finalized animations, in other words, in order to tell a story (be it a game or movie), the artist wants to express emotion in an animation, and create believable characters. Most procedural animations techniques however focus on \emph{plausibility} and do not provide the animation artists with the necessary tools to change the behavior of the synthesized animation. Therefore, it is not surprising then that procedural animation techniques face an uphill battle to find acceptance in the animation artist community.

To create procedural animations that better fit the needs of to the artistic community, it is important to work on the timings and spacings of the animation. Most animation artists use time sheets to define the extreme and in-between poses of the animation, and to refine the handcrafted animations from this starting point. Often, an artist will spend several hours to get just one pose exactly right. The timings and spacings are also used to create the illusion that an animated character is a lightweight character, with for example an energetic walk or a  voluminous character, with a slower gait and more arching movements. 

With these considerations in mind, it was clear that a procedural animation system moves  away from black-box solutions and presents the artists with an array of parameters that have a significant impact on the timings and spacing of the synthesized motions. The system must also present these parameters in a language that is easy to grasp for an animation artist. An other consideration was that these same parameters should be defined in a way that enables a machine learning approach. A system with these characteristics can learn from handcrafted animations or motion capture data, this way enabling the animation artist to convert nonprocedural animations into a procedural version. In doing so, the style of the source animation is copied into the parametric model.  The remainder of the paper is as follows. In Section \ref{section:previouswork} related work is discussed. In Section \ref{section:Animation} the architecture of our new approach to procedural animation is described.  Finally in Section \ref{sec:conclusions}, the results and future directions are discussed.

\section{Related Work}
\label{section:previouswork}
A first approach for procedural animation is \emph{animation re-purposing}. This technique uses an existing (prerecorded or handcrafted) animation and re-targets the animation for a different purpose. For example, if a character climbs a ladder, the animation is “re-purposed” to suit different widths and step sizes of the ladder. Animation re-purposing can be implemented using inverse kinematics \cite{rune2009} to adapt existing motion capture data or handcrafted animations to new environments. A downside of this technique is that there is no method to define the timings of the produced animations. Another drawback is that one or more basic animations still have to be produced.

The solution by Michiel van de Panne et. al\cite{vandepanne2000} utilizes a \emph{real time physics simulation} that creates an animation by exerting forces on the limbs of the articulated skeleton.  Although this approach produces good visual results for human characters (although with a slightly robotic feel), it is not feasible for an animation artist to adapt the synthesized motions for his own needs. A commercial solution that takes this concept further is the Endorphin commercial application.

In the gaming industry \emph{blending}\cite{ikvsblending} is commonly used to create a wide range of animations from a smaller set of handcrafted animations. This way, for example, the animation artist provides animations for a character walking on surfaces with different inclinations (e.g. starting from -20 degrees to 20 degrees in increments of 4). The disadvantage here is that this increases the workload on the animation artist, and even for a production with a large budget, this is a resource intensive solution in terms of man-hours and introduces a large amount of additional testing to verify that all the blend state transitions do not produce artifacts. 

\emph{Motion graphs}\cite{Kovar:2008:MG:1401132.1401202,chen2011} are another method to produce animations from motion capture data. Motion\linebreak graphs extract samples from motion capture data and merge them together after the path planning stage to form the animations for use in a game. Although this approach is capable of capturing the style of the human actor, the generated animations are over-fitted towards the animation from the motion capture data.

Finally, \emph{Hidden markov models}  together with a machine learning technique\cite{Tilmanne6079371} have been used to learn the style (i.e. emotion) of (level) walking of a human character. The generated animations are a blend of motion captured animations, but the generated animations can not adapt to inclined surfaces. 

\section{Animation System Overview}
\label{section:Animation}

The main difference with previous work is that this paper proposes a system that is parametrized in a way that animation artists can understand, and that also allows the procedural system to learn from motion capture data. The system is therefore not a black box system and gives the animation artist more fine grained control over the final output of the synthesized animation. Furthermore, the system is based on control theory and this allows the synthesized animations to take the environment into account.

To synthesize animations an articulated skeleton\linebreak (i.e. a rig) is used with eight degrees of freedom (hip, knee, ankle and ball of the foot of each leg)\cite{aggarwal97}. For a full simulation however eight degrees of freedom \emph{per leg} would be required \cite{Sias262401}. 

The animation system receives the current rotation angles of the skeleton and the rotations towards the target of the animation (in the case of walk animations, this is the target location of the foot), and outputs the angular velocities for the joints of the skeleton. The angular velocities are calculated with the help of a fuzzy controller (FC). One advantage is that it is possible to manually define smooth functions based on one or two input variables. A second advantage is that techniques\cite{Berenji159061} exist to convert a fuzzy controller to a fuzzy neural network (FNN). The parameters of the originating fuzzy controller are visible as weights in the FNN, which makes it possible to adapt the parameters of the animation after the network was trained with motion capture data.

\begin{figure}
\centering
\def\svgwidth{0.4\textwidth}
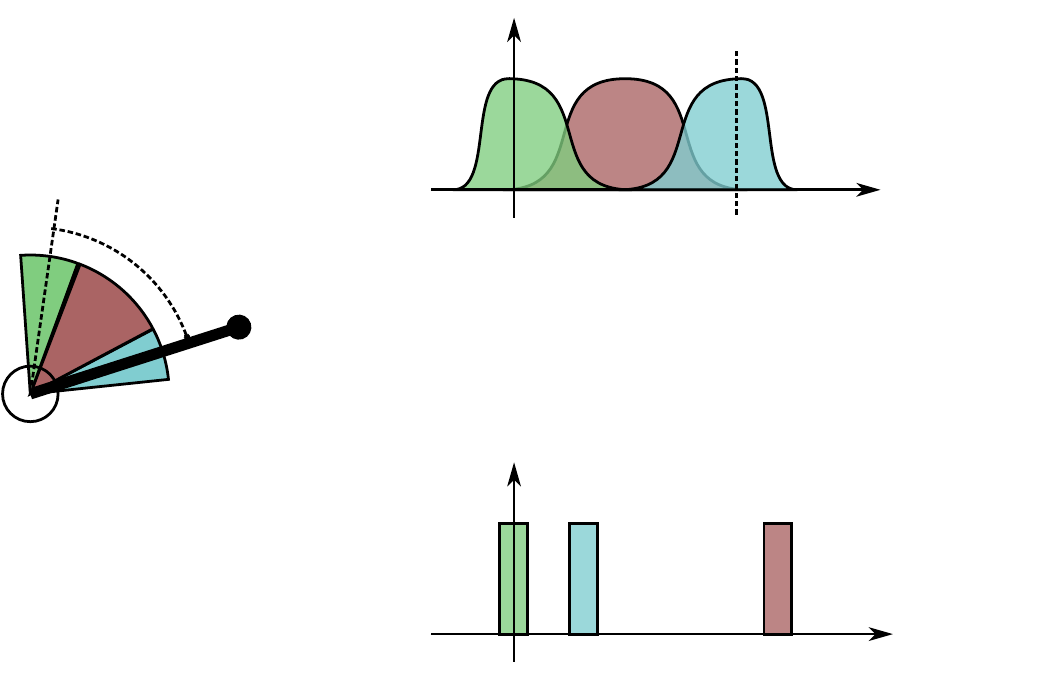
\caption{Fuzzy logic control system for smooth movement of a limb.}
\label{fig:fuzzylogic}
\end{figure}

For example, if a limb is required to rotate smoothly towards a certain target position, a fuzzy logic variable $\alpha$ (the angle between the current position of the limb and the target) can be constructed with the membership functions start, center and end. It is physically impossible to start moving at a large rotation speed. Therefore, if $\alpha$ belongs to the start set of the fuzzy variable, the output fuzzy variable is set to slow. Likewise, if $\alpha$ belongs to the center or end set, the output fuzzy variable is set to fast and stay respectively. This fuzzy control system is shown in Figure \ref{fig:fuzzylogic}.

\begin{figure}
\centering
\def\svgwidth{0.35\textwidth}
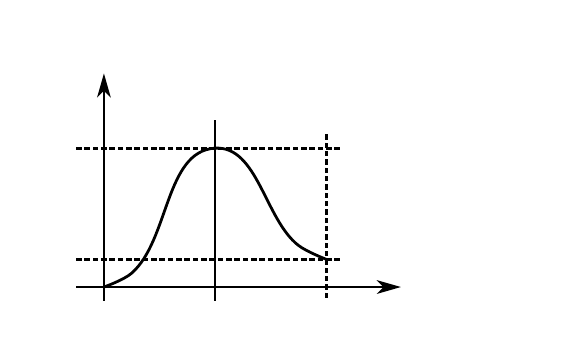
\caption{Rotation speed in function of the fuzzy variable $\alpha$.}
\label{fig:fuzzylogic2}
\end{figure}

This simple set of rules and fuzzy variables, creates a smooth input/output relationship between the variable $\alpha$ and the output velocity. When the first derivative of the output velocity is zero (or very small) at the start and end positions, the generated animation has an ease in / ease out\cite{williams2009} character.  With simple rules and membership functions it is possible to create complex decision surfaces or functions, as shown in Figure~\ref{fig:fuzzylogic2}.

\section{Controller Design}
\label{sec:controllerdesign}

There are many approaches to define the rules that make up the controller system. However with the stated goal of an FNN that can learn from motion capture, there are a couple of considerations that need to be observed. The first consideration is that all the parameters for the animation must be defined in the fuzzy controller, either explicitly by defining a fuzzy rule, or implicitly by fine tuning the membership functions and output angular velocities. This constraint ensures that motion capture data can be used to train the parameters of the controller. Furthermore, the number of membership functions for each fuzzy variable has to be as low as possible. Because we are creating controllers for the sagittal plane only, this constraint is essential to create a system that is easily extensible. As a final consideration, the rules must be robust, in other words, the controllers must be able to cope or recover from awkward start situations. 

\begin{figure}
\subfloat[Definition of $\delta$ as metric for level walking] {
\label{fig:controllerprinciple}
\def\svgwidth{0.2\textwidth}
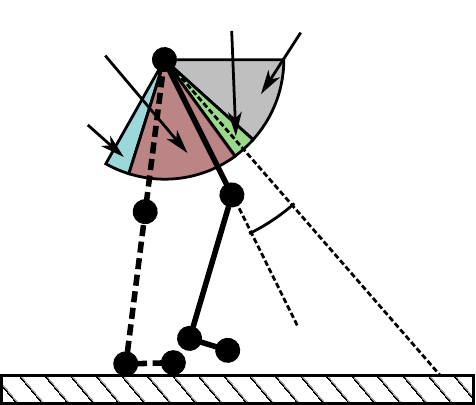
}
\hspace{5px}
\subfloat[Definition of $\delta$ as metric for ascent] {
\def\svgwidth{0.2\textwidth}
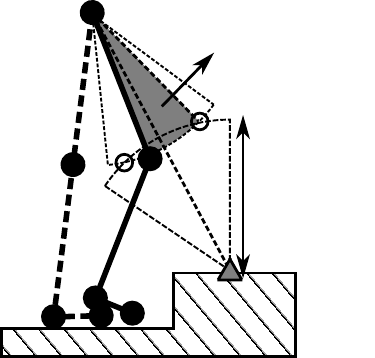
\label{fig:controllerprinciple2}
}
\label{fig:controller}
\caption{Angle $\delta$ as metric for joint rotation  }

\end{figure}

With these considerations in mind, it was important to define one or at most two metrics, for each joint in the articulated skeleton. Some metrics ware easily found. For example, if the foot is placed on the ground, the ankle joint will rotate based on the angle between the sole the foot and the surface. However, the ankle joint will only rotate when the leg is swinging forward, thus a second metric is necessary to determine the timing of the rotation of the ankle. Experience learned that the metric that provides the best results was the angle of the upper leg with the hip-target segment  (parameter $\delta$ in Figure \ref{fig:controllerprinciple}). 

This angle is not adequate however when the knee is flexed at the end pose (for example in the case of ascending a stair). In that case, the angle $\delta$ is determined by calculating the end position of the knee (as defined in Figure \ref{fig:controllerprinciple2}). Because the end position is dependent on the location of the hip joint, the calculated intersection is not a constant. However this does not pose a problem as the hip joint rotates monotonically towards its end location.

\begin{figure}
\centering
\def\svgwidth{0.3\textwidth}
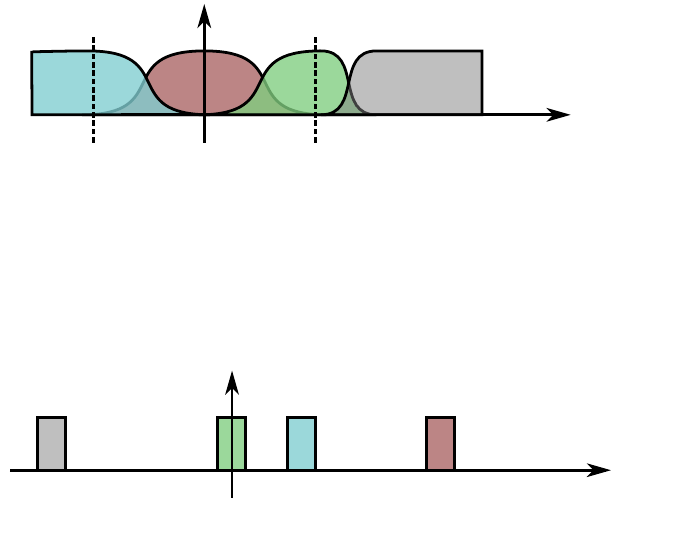
\caption{Upper leg control with ease-in ease-out behavior.}
\label{fig:hiprules}
\end{figure}

In a similar manner as in Figure~\ref{fig:fuzzylogic} rules can be defined that will guide the upper leg in swing phase towards the desired location, with a smooth input output relationship between the variable $\delta$ and the output velocity for the hip joint (Figure \ref{fig:hiprules}). Because the absolute value of $\delta$ can change however, and the membership functions are constant, a scaling operation is performed on this parameter. For level walking, the parameter $\delta$ is mapped to \emph{one} if the upper leg has arrived at the required location, and to \emph{minus one} if the upper leg is at its start position. The logical choice is then to map $\delta$ to \emph{zero} when the upper leg has a zero rotation angle relative to the hip.
\emph{Ascending} a stair or steep terrain has other dynamics then walking on flat terrain. It is however possible to use the same angle metric $\delta$  as defined in Figure \ref{fig:controller}. Except for the leg in stance phase, the rules are now very similar to the rules that were defined for level walking.
\balance
\section{Results}
\label{sec:results}
\begin{figure}
\centering
\tiny
\def\svgwidth{0.25\textwidth}
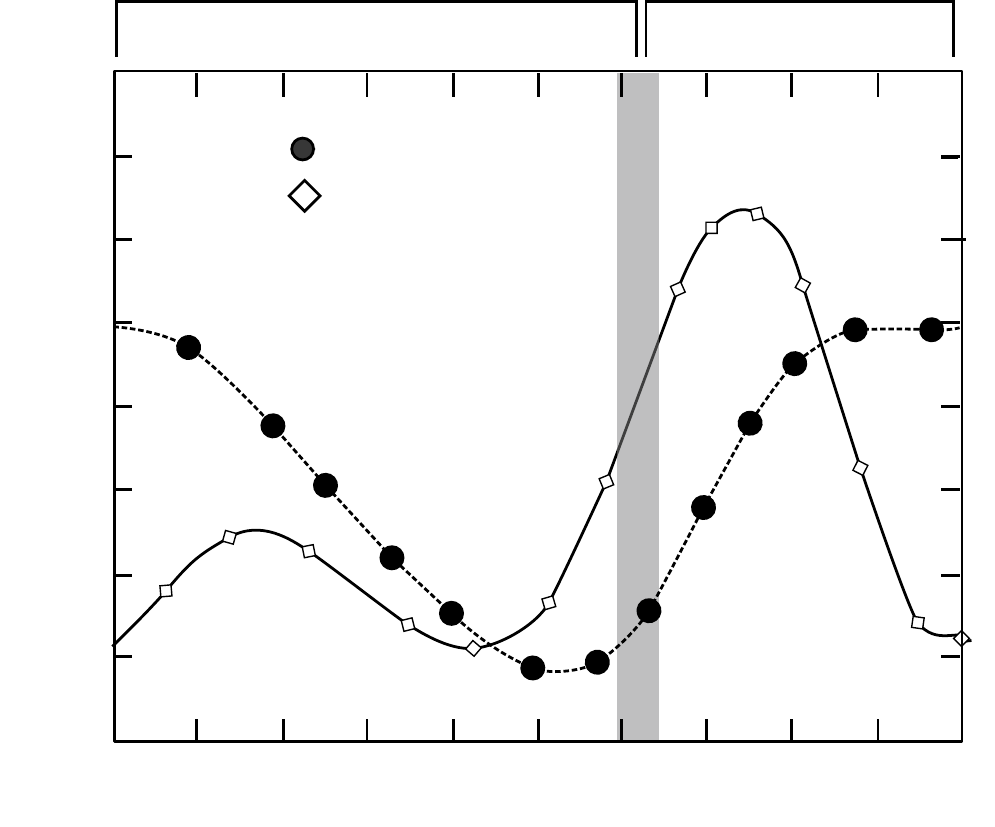
\caption{Knee and hip joint reference rotation curves for level walking.\cite{Riener200232,Hansen2004807}}
\label{fig:jointcurves}
\end{figure}

\begin{figure}
\centering
\subfloat[level walk, stepsize  60cm] {\def\svgwidth{0.22\textwidth}
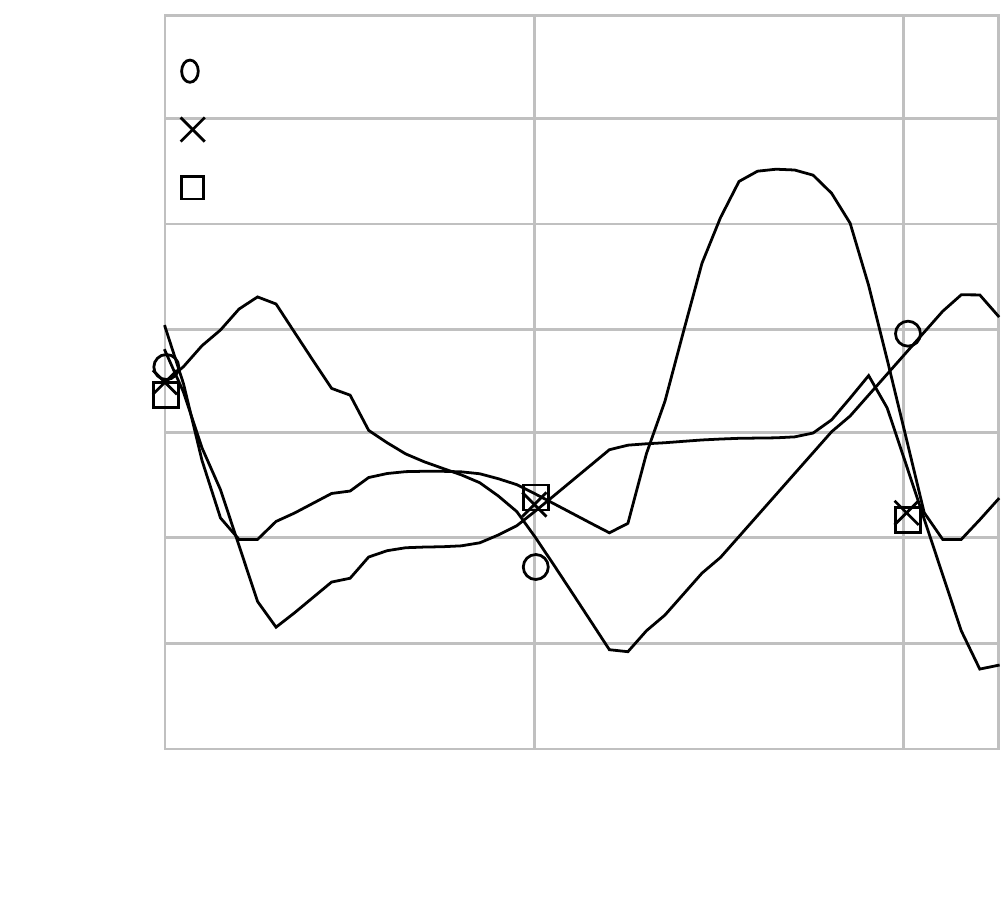}
\vspace{5px}
\subfloat[level walk, stepsize 70cm] {\def\svgwidth{0.22\textwidth}
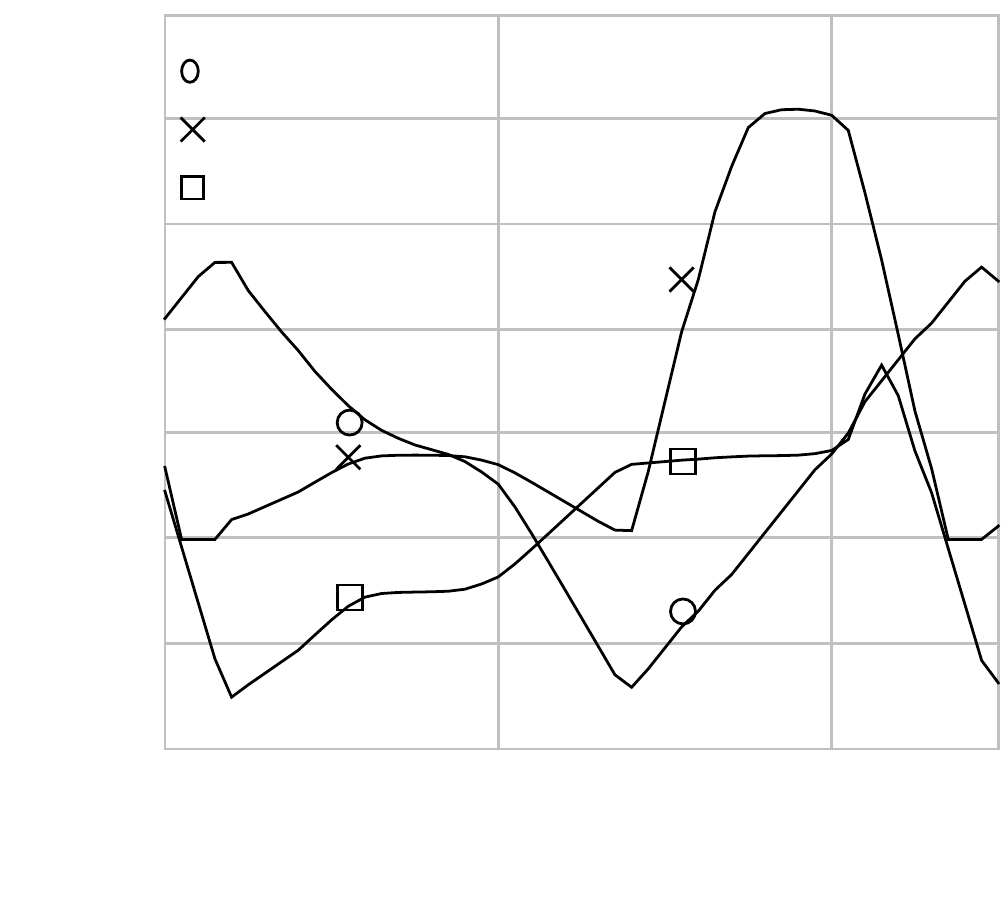}
\vspace{5px}
\caption{Procedural joint angle curves for level walking for various step sizes.}
\label{fig:levelwalking}
\end{figure}

The reference curves for level walking are given in Figure \ref{fig:jointcurves}. The rotational curves for the real time generated level walking gait cycles are given in Figure~\ref{fig:levelwalking} for two different step sizes.

Video footage illustrating the human gait on flat surface and when walking a stair can be found on \url{http://vimeo.com/user10476935/channels}.  The video fragment illustrates that it is possible to handtune fuzzy controllers for level walking and ascending, for a human (biped) character. 
\section{Conclusion And Future Work}
\label{sec:conclusions}
In this paper it was shown that it is possible to create procedural animations by manually tuning fuzzy controllers. However,
to create convincing and realistic animations it is necessary to tune the fuzzy controllers with a machine learning approach, and to engage animation artists that can help identify problem areas in the generated animations. Another future machine learning experiment would try to discover relationships between the input variables for the articulated skeleton. Finally, the system will extended to a full body simulation and more basic animations (such as running, jumping and swimming).
\balance
\nocite{Anonymous:2012}
\bibliographystyle{spmpsci}
\bibliography{cgishort}

\end{document}